\documentstyle[aps,prb]{revtex} 
\draft
\begin{document}

\title{Polarizability of 2D monster and light scattering}

\author{M.V. Entin and G.M. Entin}

\address{
Institute of Semiconductor Physics, Siberian Branch, Russian
Academy of Sciences.  \\
13 Lavrent'ev st., Novosibirsk, Russia 630090.
E-mail: entin@isp.nsc.ru}

\maketitle

\begin{abstract}
The electrostatics of 2D system with complicated inner
boundary is studied. The object which we call "monster" is built by
an iterative process of multiple conformal mapping of the circle
exterior.   The procedure leads to the figures
built from circle with branching curved cuts (i),  to
the multiple tangential near-round circles (ii) and the
Mandelbrot map (iii). The polarizability of a monster in the
homogeneous external field is found.  The light scattering
cross-section was expressed through the polarizability of a
monster.
\end{abstract}




{\bf Introduction.}
The fractal aggregates appear in many experimental and theoretical
situations.  Such aggregates are the topic of the diffusion limited
aggregation, the theory of percolation and the electric breakdown.

The purpose of the present paper is to study the polarizability of
solitary particles with complicated boundaries. We shall refer the
complicated particles, built from small subunits, as ''monsters'',
distinguishing them from fractals, which result from the infinite
growth. The monsters can be obtained both directly in the growth
process, and as a result of Van der Waals aggregation of
preproduced spherical nanoclusters into large snowballs. It is
natural to expect that in many real situations  medium is
composed from the monstrous or fractal objects, rarely embedded
into matrix.

The optical properties of fractals, including  polarizability  were
studied in Ref.
\CITE{Stroud,Claro,Brouers,Sh,scale,Bess,ShalaevBotet,d,d1}.
The attention of these studies was mostly concentrated on the
localization of high-frequency polar eigenmodes of fractals and
corresponding large fluctuations of local electric fields.

The works,\cite{Sh,d,d1} most close to the purpose of the present
study, consider the fractal as a molecule built from coupled
dipolar monomers. The system of linear equations for dipoles was
solved either in self-consistent approximation \cite{Sh} or by
computer simulations. \cite{d,d1}

The self-consistent approximation is valid only if the polarizability
of the fractal medium is not large. The large polarizability of the
medium leads to the strong screening of the external field. This is
the case for metal clusters, studed in the present work.

Unlike the referred papers, we are interested in the situation
where the local fields are strongly deviate from the mean value. We
shall consider the aggregate interior as a macroscopic medium,
neglecting the atomic structure effects.  In this approach one
should solve the macroscopic Maxwell equation inside and outside of
fractal aggregate with some boundary conditions on its surface. We
shall deal with aggregates, that are small in comparison with the
wavelength. This permits to solve the Laplace equation, instead of
the Maxwell one.

The Laplace problem is significally simplified, if one does not
need the solution both inside and outside of the object. It is so,
if the dielectric permittivity of the cluster is essentially
larger or essentially smaller than that of the external medium. If
the cluster medium is metal, then its permittivity is determined by
the formula $1-\omega_p^2/\omega^2$, where $\omega$ is the light
frequency, and $\omega_p$ is the plasma frequency. The first case
corresponds to $\omega\ll \omega_p$, the second one to the
frequency near the bulk plasma frequency.

We shall study the 2D case, which allows to use the theory of
analytical functions for solution of the Laplace equation. This
simplification may be considered as a model for the 3D case. At the
same time, it is applicable to the clusters with cylindrical shape
$F(x,y)=\mbox{const}$.

We shall find the electric polarizability of a large aggregate of
complicated shape that can be constructed by means of multiple
conformal maps. In this case the exact solution of the problem can
be found.  One kind of such construction of 2D fractal objects was
suggested in Ref. \CITE{Has,Proc,DHO}. The fractal was treated as
a result of iterative mapping of a simple domain, for example,
exterior of a circle. The elementary step maps the exterior of the
circle onto the exterior of a circle with a bump or with a strike.
The multiple mapping complicates the form of the domain, producing
a bump (strike) for a step of an iteration. This construction
permits to simulate growth, guided by so called harmonic measure,
when the addition of an element of fixed area to the domain appears
with the probability, proportional to the diffusion flow,
determined by the solution of the harmonic equation of the
diffusion (DLA).

Another approach to constructing fractal is a simple map, for
example, square map \cite{Mandel} $z_{n+1}=c-z_n^2$. The multiple
superposition of this map separates the complex plane into the
repulsion basin, composed from all points $z_0$ for which $z_n\to
\infty$. The boundary of that basin is a fractal, called Julia set.

The limited repetition of maping produces a monster, the complicated 
but not fractal object. We are interested in the properties of the 
monster as well as the properties of the limit of the infinite number 
of repetitions, the fractal.  The complexity of the fractal grows 
linear with the map order. It differs this construction from the 
square map of Mandelbrot, where the complexity grows exponentially 
with the map order.

In the case  of rare inclusions the effective electrical properties of
medium are determined by individual monster properties. For
example, the effective dielectric permittivity of the medium,
containing  monsters with concentration $n$ and polarizability
$\chi_{ij}$ in the limit of low concentration is
\begin{equation} \label{1}
\epsilon_{ij} = \delta_{ij} +
4 \pi n \chi_{ij} .
\end{equation}
Similar formula   $\sigma_{ij} = \sigma(\delta_{ij} +
4 \pi n \chi_{ij})$ determines
the effective conductivity of the medium
with conductivity $\sigma$, containing insulation inclusions.

The other important property of physical structures is the light
scattering cross-section. It is determined by the polarizability,
if the wavelength exceeds the diameter of the object.
 In particular, we shall deal with the polarizability and
with the scattering of light on a small metal particle.

{\bf Method.}
These
properties can be found by solving the Laplace equation
$\phi$ ~ $\Delta\phi({\bf r})=0$ for  the potential outside of a
solitary particle in the external uniform electric field
$E=E_x+iE_y$.  For  $\omega\ll\omega_p$ the boundary conditions at
the particle surface $S$  and at the infinity are \begin{equation}
\partial_S\phi=0, ~~~ \phi|_\infty=-Re(E^*w).  \end{equation}
\noindent Here $w=x+iy$.

The effective conductivity of media with rare
dielectric inclusions is determined by the solution of the
current flow problem around the inclusion.

If $\omega-\omega_p\ll\omega_p$,  the
potential  $\psi$ satisfies the condition
$\partial_{\bf n}\psi|_S=0$ at the boundary of the inclusion.
According to the Couchi-Rieman conditions $\phi$ and $\psi$ are the
real and imaginary parts of the complex potential $\Phi$.
This case corresponds to the problem of the current flow
in a conducting medium around  an insulating inclusion with the
same shape.  

The idea of the present work is to construct a complicated
object from simple  one by means of repetition of some
conformal map $f(z)$. The map should transform the exterior of the
basic object to the exterior of complicated one and does not move
the infinity point.  It doesn't alter the Laplace equation and the
surface boundary conditions (with the accuracy to the coefficients).
Therefore, the solution of the problem for the source object gives
it for the image object too.

Let us consider a multiple map $F^{(n)}$ composed of the n basic
maps $f_\lambda (z)$, where $\lambda$ is a parameter (or a set of
parameters) of the map, and some additional map $G$:
\begin{equation} \label{eqFn}
F^{(n)}(w)=G\raisebox{0.5ex}{\scriptsize o}
f_{\lambda_1}\raisebox{0.5ex}{\scriptsize o}
f_{\lambda_2}\raisebox{0.5ex}{\scriptsize o}...
\raisebox{0.5ex}{\scriptsize o}
f_{\lambda_n}(w).
\end{equation}
The parameters $\lambda_i$ may be chosen fixed, random, or depended
regulary on $i$.

If we are interested in the dipole component of electric field only,
we should take into account only $f_1$ and $f_{-1}$ coefficients of
series expansion of the map
$f_{\lambda_i}(w)=f_1^{(i)}w+f_0^{(i)}+f_{-1}^{(i)}w^{-1}+...$.

The complex potential outside the unite circle in the external
uniform electric field in 2D case is \begin{equation} \label{eqPhi}
\Phi(w)=-(E^* w -E/w), \end{equation} on the complex $w$-plane. Let
$G\equiv 1$.  Substitution of the series expansion of the  inverse
function $w(z):  F^{(n)}(w(z))\equiv z$ into (\ref{eqPhi}) gives us
the polarizability tensor $\chi$:
\begin{eqnarray}
\label{eqBumpChi}
\chi=\frac{A}{2}
\left(
  \begin{array}{cc}
  Re(B)-\sigma    &    Im(B)  \\
  Im(B)           &   -Re(B)-\sigma
  \end{array}
\right),
\end{eqnarray}
\noindent
where $\sigma=-1$ in the case of metal particle and $1$ in the
case of dieletric one. $A$ and $B$ are defined as follows:
\begin{eqnarray}
\label{eqBumpTerm2}
A&=&\left|\prod_{i=1}^n f_1^{(i)}\right|^2,   \\
B&=&\frac{1}{A}\left[- f^{(1)}_1 f^{(1)}_{-1} +
f^{(n)}_1 \sum_{k=2}^n f^{(k)}_{-1} \prod_{i=1}^{k-1} \left(f^{(i)}_1
\right)^2 \right].
\end{eqnarray}
The principal values of polarizability tensor are
$$\chi_{1,2}=-\frac{A}{2}(\sigma \pm |B|)$$

\begin{figure}[b!] 
\centerline{
\input epsf
\epsfysize=4cm
\epsfbox{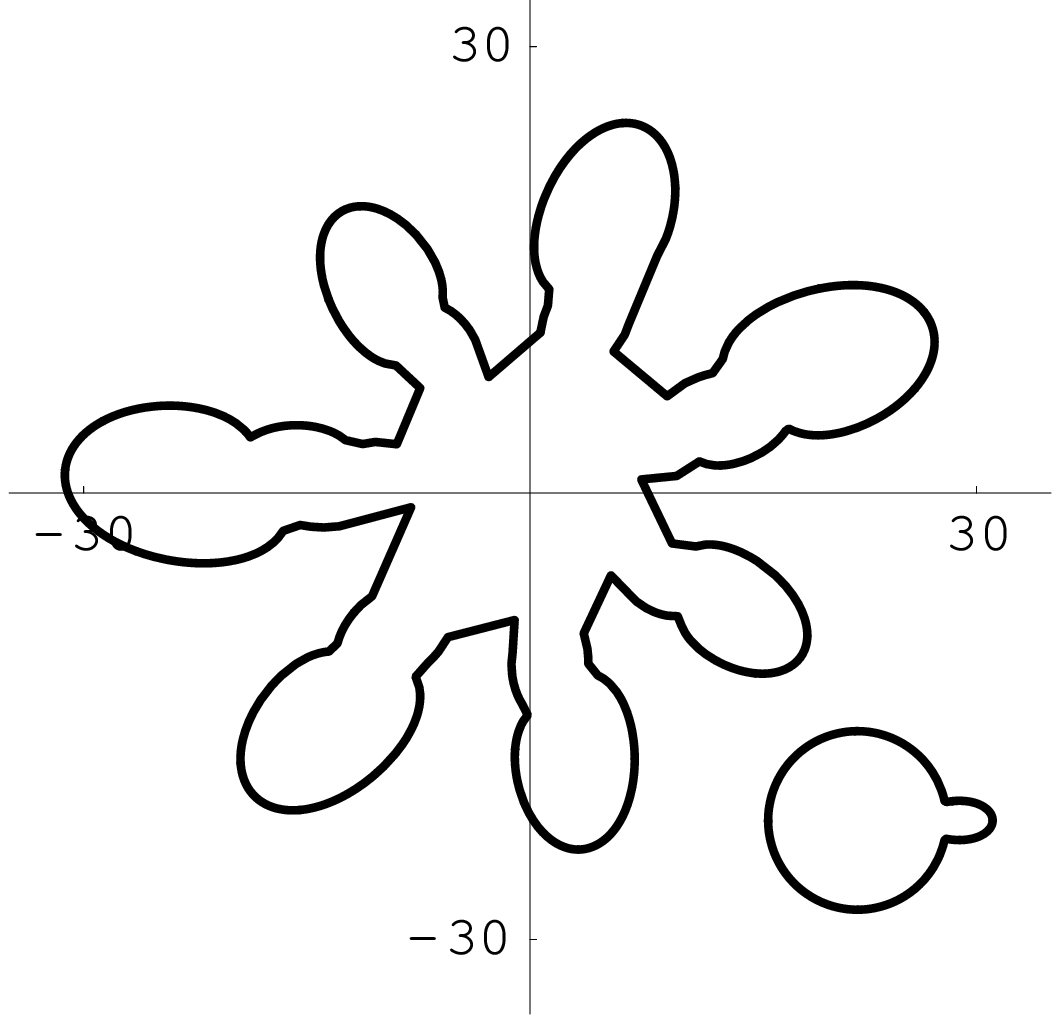}~~
\epsfysize=4cm
\epsfbox{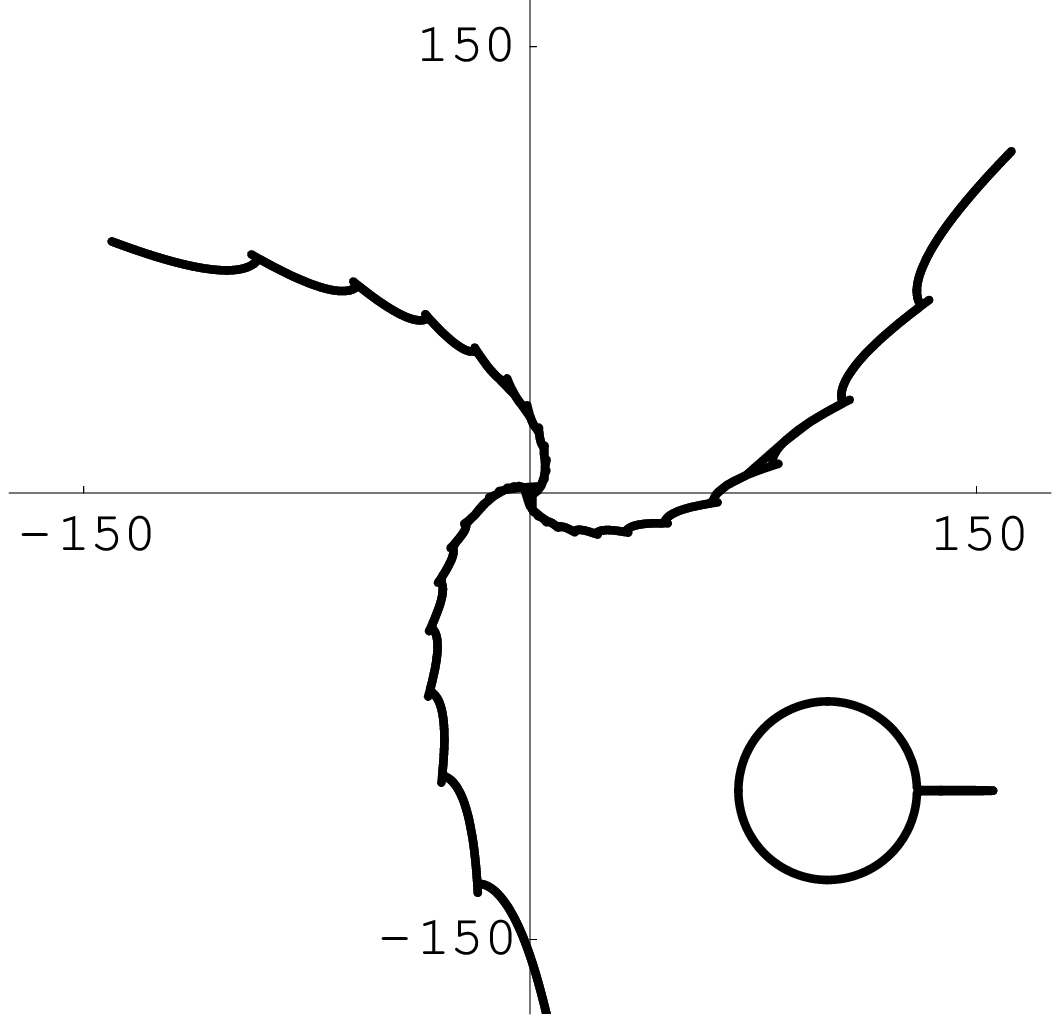}
}
\vspace{10pt}
\caption{
The case (i). The images of the unit circle according to the
Eq. \ref{eqBumpMap} after $n=50$ iterations for parameters
$\theta=2.7$, $a=2/3$, $\lambda=0.1$  (case of strikes, figure a)
and $\theta=2.15$, $a=1$, $\lambda=0.1$ (case of bumps, figure b).
Images of basic maps applied to the unit circle are shown in
the corners (the size of basic map is arbitrary).
} \label{fig1}
\end{figure}

{\bf Examples.}
In the recent works \cite{Has,Proc,DHO} a multiple conformal map
(i) was used for constructing of models of fractal objects, in
particular, DLA and dielectric breakdown clusters. The map
(\ref{eqBumpMap}) transforms the unit circle to the unite circle
with a small bump (or strike) on it (Fig.~\ref{fig1}).
\begin{eqnarray}
\label{eqBumpMap}
\nonumber
f_{\lambda}(w)&=&w^{1-a}\Bigg\{\frac{1+\lambda}{2w}(1+w)\Bigg[1+w+\\
&&+w\left(1+\frac{1}{w^2}-\frac{2}{w}\frac{1-\lambda}{1+\lambda}\right)^{1/2}\Bigg]-1\Bigg\}^a ,\\
\label{eqBumpMap1}
f_{\lambda,\theta}(w)&=&e^{i\theta} f_\lambda(e^{-i\theta} w).
\end{eqnarray}

The shape of the bump is determined by two parameters
$a$ and $\lambda$.  In this work we assume them to be constant. We
use the rotated map (\ref{eqBumpMap1}) as the basic for $F$ and
consider simple dependence of the angle of rotation $\theta_k=k
\theta$, where $\theta=\mbox{const}$ is the angle incommensurable
to $\pi$.  This choice enables us to find the exact equation for
the polarizability but still produces the complicated objects.
The map, recursively applied to the source unit circle,
produces monsters as on the Fig.~\ref{fig1}.

The coefficients $A$, $B$ of the polarizability tensor in the limit
$n\to \infty$ are:
\begin{eqnarray} \nonumber
A&=&(1+\lambda)^{2an}, \\
B&=&a\lambda \frac{2+(2a-1)\lambda}{(1+\lambda)^2} 
\frac{e^{2i\theta}-\beta}{\beta^2-2\beta\cos 2\theta +1} 
e^{i n\theta} 
\end{eqnarray}
\noindent
where $\beta=(1+\lambda)^{2a}$.
\noindent

\begin{figure}[b!] 
\centerline{
\input epsf
\epsfysize=5cm
\epsfbox{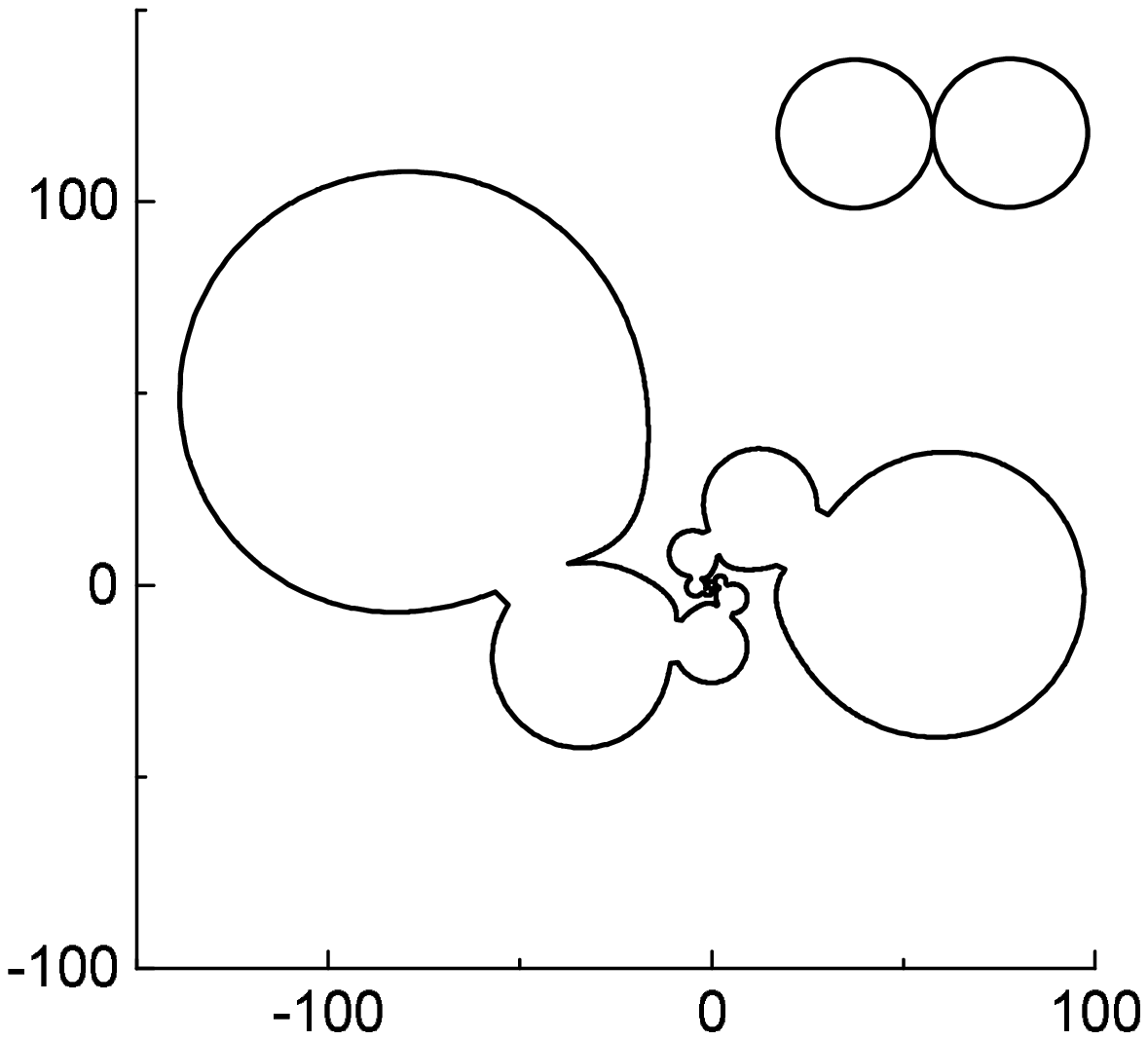}}
\vspace{10pt}
\caption{
The case (ii). The image of the circle after 10-th map
(\ref{eqCircMap}) for $\theta_n = n \theta$, $\theta=-2.1$. The
basic map is shown also.
} \label{fig2}
\end{figure}

Another considered basic map (ii) is the map of the exterior of the
unit circle to the exterior of two tangenting unit circles:
\begin{equation}
\label{eqCircMap}
f(w) = \frac{i \pi}{\log \left[ \frac{w-1}{w+1} \right]}+1 , ~~~
f_\theta(w)=e^{i\theta} f(e^{-i\theta} w).
\end{equation}
\noindent
The example of multiple map is shown on Fig.~\ref{fig2} in the case
of $\theta_k=k \theta$.

In the case of $n\to \infty$ corresponding quantities $A$ and $B$
for polarizability (\ref{eqBumpChi}) are
\begin{eqnarray}
A=2\left(\frac{\pi}{2}\right)^{2n}, &~~~& B=\frac{\pi^2
(-e^{2i\theta})^{n+1}}{3 (4+ \pi^2 e^{2i\theta})}.
\end{eqnarray}

\begin{figure}[b!] 
\centerline{
\input epsf
\epsfxsize=6cm
\epsfbox{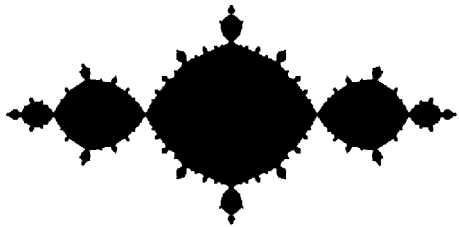}}
\vspace{10pt}
\caption{
The case (iii). The Mandelbrot fractal, obtained by the map
(\ref{eqMandMap}) of the unite circle.
} \label{fig3}
\end{figure}

The third basic map (iii) is the known Mandelbrot map. Resulting
transform is
\begin{equation}
F^{(n)}(w)=(1-(1-(...-w^2)^2...)^2)^{2^{-n}}.
\label{eqMandMap}
\end{equation}
\noindent
The additional transform $G(\alpha)\equiv\alpha^{2^{-n}}$  makes
the final map single-valued on the exterior. On Fig.~\ref{fig3} the
domain is shown, which boundary, so-called Julia set, is the image
of a unit circle.  The multiple map (iii) produces fractal, which
appears, for example, in the hierarchical  description of the
percolation transition in the complex plane of percolation
probability.\cite{scale}

The polarizability of the metallic Mandelbrot monster is
\begin{equation}
\chi_{xx,yy}=\pm\frac{1}{2}+
\left(\frac{(1+\sqrt 5)}{2}\right)^{2^{-n+1}}, ~~~ \chi_{xy,yx}=0.
\end{equation}

{\bf Light scattering.}
The 2D problem may be considered as a model for the more realistic
3D system. From the other hand, it describes the properties of
the cylindrical objects, which size in z-direction exceeds the
maximal diameter in (x,y) plane.

\begin{figure}[b!] 
\centerline{
\input epsf
\epsfysize=4cm
\epsfbox{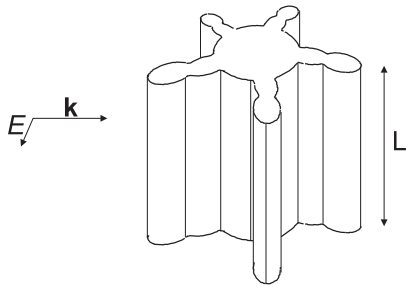}}
\vspace{10pt}
\caption{
The geometry of the light scattering problem on a
cylindrical object with complicated section. ${\bf k}$ is the wave
vector.  The plane of polarization coincide with the cross-section
plane.
} \label{fig4}
\end{figure}

In relation to the problem of light scattering we shall consider
the geometry, depicted on the Fig.~\ref{fig4}. In this case, if
the electric vector ${\bf E}=|{\bf E}|{\bf e}$ lyes in the face
plane of the monster, the cross-section is
\begin{equation}
\sigma=\frac{8\pi}{3}k^4|\chi_{ij}e_j|^2 4\frac{\sin^2
(\frac{k_zL}{2})}{k_z^2},  \end{equation}

\noindent
where $e_i$ is the vector of polarization of the electric field.

{\bf Discussion.}
The polarizability of the round metal particle in 2D is roughly
proportional to the second power of its radius. For a strongly
extended asymmetric particle the components of polarizability
tensor are determined by the sizes of the particle in the direction
of the field.

In general, the monsters in the cases (i) and (ii) have no
symmetry, so each component of tensor $\chi$ is finite. In the
case (iii) the fractal has x and y axes of symmetry  resulting in
$\chi_{xy}=0$. The monsters (i) and (ii) grow
exponentially with the number of iteration. This leads to  the 
same growth of the polarizability.

In the case (iii) the sizes of monster are limited,
as is the polarizability tensor. The limiting ratio of its
components roughly corresponds to the ratio of sizes. The fractal
structure of this system manifests itself in the exponential law of
approaching to the limit.

The purpose of the present work was to find the polarizability of
metal clusters. However, these results are more widely applicable.
They describe the effective conductivity of a medium with
insulating or perfectly conducting inclusions. In accord with the
Einstein relation, the formulae similar to (\ref{1})
determine the effective diffusion coefficient of some particles in
the medium with inclusions, non-penetrable for diffusing particles.

We have used some ways of constructing monsters and fractals by 
means of multiple conformal maps. The basic map may be more simple 
as  the Mandelbrote square map (iii) or more 
complicated as the bump map (i) or the tangental circle map (ii). The 
advantage of complicated maps (i) and (ii) is that they  permit to 
adjoin unites of fixed form to a cluster, the disadvantage is that 
the complexity of monster grows linearily with the iteration number, 
while the complexity of square map (iii) (and  more general one,  
$z_{n+1}=c-z_n^2$) grows exponentially. 

We have 
considered the sequent maps with fixed parameters only. In the cases 
(i) and (ii) the external parts of monsters are essentially 
larger than internal ones. The fine details of the monster are inside 
of it and the exterior of the monster is not fractal.  The 
polarizability of the monster depends generally on its external 
shape, scaling like the square of its diameter.  
This differs situation from one, considered in 
Ref. \CITE{Has,Proc,DHO}, where the fractal shape of a growing 
cluster is achieved by the selection of the map parameters 
$\lambda_i$ in such a way, that adjoining bumps have equal sizes.  
Our results may be easily generalized to this case also and to the 
other  cases of random, or some regular dependence of parameters on 
the map order. 
 
The specific shape of monsters (i) is similar to the branching traces 
of electric breakdown and  hence is applicable for description of 
electric properties of medium after electric breakdown.  The monsters 
(ii) have the shape of coupled droplets in emulsion. By selection of 
parameters this case may be generalized to desribe clusters of near 
round droplets with equal sizes. 

{\bf Acknowledgments.} We are thankful to Dr. E.M. Baskin
and Dr. L.S. Braginsky for helpful discussions. The work was
partially supported by the Russian Foundation for Basic Researches
(Grant 97-02-18397) and the Program "Physics of Solid
Nanostructures" of the Ministry of Science of Russian Federation.

\end{document}